# Barkhausen effect in the first order structural phase transition in type-II Weyl semimetal MoTe$_2$

Chuanwu Cao[1,2], Xin Liu[1,2], Xiao Ren[1,2], Xianzhe Zeng[1,2], Kenan Zhang[2,3], Dong Sun[1,2], Shuyun Zhou[2,3], Yang Wu[4], Yuan Li[1,2], Jian-Hao Chen*[1,2]

[1]International Center for Quantum Materials, School of Physics, Peking University, NO. 5 Yiheyuan Road, Beijing, 100871, China
[2]Collaborative Innovation Center of Quantum Matter, Beijing 100871, China
[3]State Key Laboratory of Low Dimensional Quantum Physics, Department of Physics, Tsinghua University, Beijing 100084, China
[4]Tsinghua-Foxconn Nanotechnology Research Center, Tsinghua University, Beijing 100084, China.

E-mail: chenjianhao@pku.edu.cn

Abstract:

We report the first observation of the non-magnetic Barkhausen effect in van der Waals layered crystals, specifically, between the $T_d$ and $1T'$ phases in type-II Weyl semimetal MoTe$_2$. Thinning down the MoTe$_2$ crystal from bulk material to about 25nm results in a drastic strengthening of the hysteresis in the phase transition, with the difference in critical temperature increasing from ~40K to more than 300K. The Barkhausen effect appears for thin samples and the temperature range of the Barkhausen zone grows approximately linearly with reducing sample thickness, pointing to a surface origin of the phase pinning defects. The distribution of the Barkhausen jumps shows a power law behavior, with its critical exponent $\alpha = 1.27$, in good agreement with existing scaling theory. Temperature-dependent Raman spectroscopy on MoTe$_2$ crystals of various thicknesses shows results consistent with our transport measurements.

# Introduction

Layered transition metal dichalcogenides (TMDs), a family of van der Waals crystals, have attracted extensive research interests due to their intriguing properties[1-4]. Certain TMDs have rich structural and electronic phases which exhibit drastically different physical properties[5-15]. MoTe$_2$, in particular, is an excellent example of such multi-phase materials, which possesses three different crystal structures: hexagonal 2*H*, monoclinic 1*T'*, and orthorhombic *T*$_d$[14]. The 2*H* phase MoTe$_2$ is a semiconductor and candidate material for two-dimensional field effect transistors (2D-FET)[16]. The 1*T'* phase MoTe$_2$ is a quantum spin Hall candidate at the single layer form [17,18], and a central symmetric semimetal at the multi-layer form, whose crystal symmetry belongs to the P2$_1$/m space group [14]. The non-centrosymmetric *T*$_d$ phase MoTe$_2$ belongs to the Pmn2$_1$ space group and is a type-II Weyl semimetal which violates the Lorentz invariant, with the Weyl points appear as the touching points between electron and hole pocket in a tiled cone configuration[14,15].

The structural phase transition of MoTe$_2$ between the 1*T'* phase and the 2*H* phase can be achieved by electrostatic gating at room temperature [8], while the transition between the 1*T'* phase and the *T*$_d$ phase can be achieved by varying the sample temperature[14]. The temperature driven structural phase transition between *T*$_d$ (low temperature phase) and 1*T'* (high temperature phase) has been demonstrated by electrical transport[19, 20], Raman spectroscopy [14] and angle-resolved photoemission spectroscopy [15]. In this article, we report the first observation of Barkhausen physics in the first order structural phase transition between the two metallic phases of MoTe$_2$ (the 1*T'* phase and the *T*$_d$ phase).

The Barkhausen effect is originally defined as a series of sudden reversal of Weiss domains in a ferromagnet during a continuous process of magnetization or demagnetization[21-23]. The Barkhausen physics is related to domain-wall pinning and de-pinning when a ferromagnetic system experiences a transition between two different states. During the transition, the domain wall between two states moves through the sample. Defects in the crystal pin the moving domain wall and hold the neighbor area in the past state until the energy gain in flipping the whole neighbor area gets larger than the de-pinning energy. When de-pinning happens, the pinned area suddenly turns into the other state, leading to a jump in the total magnetization of the sample (see supplementary materials figure S1). The Barkhausen effect has been a powerful tool to characterize magnetic and ferroelectric materials[24-27], and it has also attracted growing interest as an example of complex dynamical systems displaying dimension-dependent scaling behavior [22]. Given the ubiquitous presence of the Barkhausen effect in magnetic phase transition[24-29], only a limited number of research has revealed Barkhausen physics in thermally-driven first-order phase transition in non-magnetic materials[30-35], and none of such transition has been observed in van der Waals layered crystals.

Here, we report the first observation of the thickness-dependent Barkhausen effect in the thermally-driven first order structural phase transition between the $T_d$ phase and the $1T'$ phase in MoTe$_2$. The $T_d$ phase and $1T'$ phase MoTe$_2$ are both metallic but with different resistivity at any given temperature (e.g. $\rho_{T_d} < \rho_{1T'}$). Thus, careful measurement of the resistivity of the MoTe$_2$ as a function of temperature $\rho(T)$ reveals events of the structural phase transition in the crystal.

For bulk crystals, the phase transition happens sharply at around 250K with no additional features, and hysteresis of the transition temperature, defined as $T_{hysteresis} = T_{T_d \to 1T'} - T_{1T' \to T_d}$, is around 40K (figure 1b). For samples between 100nm-20nm, on the other hand, the phase transition occurs in a large temperature region and exhibits a series of kinks in $\rho(T)$, which is a manifestation of the Barkhausen effect. Thinning down the MoTe$_2$ crystal from bulk material to about 20nm results in a drastic strengthening of the hysteresis in the phase transition, with $T_{hysteresis}$ increasing from ~40K to more than 300K. The temperature range of the Barkhausen zone grows approximately linearly with reducing sample thickness as determined by four-probe resistivity measurement. The distribution of the Barkhausen jumps shows a power law behavior, with its critical exponent $\alpha = 1.27$, consistent with theoretical expectations as explained later in this letter. Temperature-dependent Raman spectroscopy is also performed on MoTe$_2$ crystals of various thicknesses, showing results consistent with our transport measurements.

## Results and discussion

The temperature-driven structural phase transition in few-layer metallic MoTe$_2$ is studied using electrical transport measurement as well as Raman spectroscopy. Figure 1a shows the atomic schematics of MoTe$_2$ in $T_d$ phase and in $1T'$ phase. The $T_d$ phase MoTe$_2$ shares the same in-plane crystal structure with the $1T'$ phase MoTe$_2$ and differs only in vertical stacking[14]. Figure1b shows a typical temperature-dependent four-probe resistivity curve of bulk metallic MoTe$_2$. The transition from $1T'$ phase to $T_d$ phase occurs at about 230K to 250K and the transition from $T_d$ phase to $1T'$ phase occurs at about 260K to 270K. It is experimentally observed that $\rho_{T_d} < \rho_{1T'}$ at the phase transition temperature, thus $\Delta\rho_{T_d \to 1T'} > 0$ and $\Delta\rho_{1T' \to T_d} < 0$.

Figure 1c shows a typical temperature-dependent four-probe resistivity curve $\rho(T)$ for a 50nm-thick MoTe$_2$ sample, which is fairly different from that of the bulk samples. The cooling curve and warming curve do not overlap from 80K to 320K, showing that the hysteresis of transition between the $1T'$ phase and the $T_d$ phase has been strengthened. While bulk crystals show a smooth function of $\rho(T)$ during the phase transition, a series of kinks in $\rho(T)$ appear in a wide temperature range for both the warming and cooling curves in thinner samples. Here a kink means an instance of rapid change in the $\rho(T)$ curve. We found that all the kinks in the cooling curve ($1T' \to T_d$, upper inset in figure 1c) represent rapid drops in the resistivity, while all the kinks in the warming curve ($T_d \to 1T'$, lower inset in figure 1c) represent rapid increase in the resistivity, consistent with the fact that $\Delta\rho_{1T' \to T_d} < 0$ and $\Delta\rho_{T_d \to 1T'} > 0$. Thus, both the kinks and the strengthened hysteresis behavior in the $\rho(T)$ curves can be attributed to modified structural phase transition in thin MoTe$_2$ crystals. The appearance of kinks in the $\rho(T)$ curves reveals the existence of phase pinning in the first order phase transition process in thin MoTe$_2$. During cooling (warming) process, defects or local strains in the crystal pin their neighbor area in the $1T'$ ($T_d$) phase; as the temperature becomes sufficiently low (high), the pinned area suddenly flips to the $T_d$ ($1T'$) phase, and be detected as a kink in the $\rho(T)$ curves.

Figure 2a shows $\rho(T)$ curves for seven temperature scans of a 50nm MoTe$_2$ sample between 10 K and 350K(curves offset for clarity). All the curves show metallic and hysteresis behavior. We found that kinks in the $\rho(T)$ curves do not happen at the same temperature during different experimental runs for the same device, but rather, has a statistical distribution, which is analogous to the Barkhausen effect observed in magnetic materials (e.g. jumps in the *M(H)* curves has a statistical distribution)[24, 25]. The distribution of the kinks can be summarized as a histogram of the occurrence of the kinks weighted by the height (Δ*ρ*) of respective kinks, as shown in figure 2b. We collect kinks from all 7 runs of cooling and warming process, sum up all kinks' height in each bin (5K in temperature), and plot the value (in units of Ohms) as a function of temperature. The height of a kink is obtained using a procedure shown in the inset of figure 2a. It is worth pointing out that the kink histogram may not show the full scale of the phase transition, but rather, show the temperature distribution of all the detected Barkhausen jumps. Nonetheless, the kink histogram reveals the temperature range that phase transition occurs. In figure 2b, the blue histogram is extracted form cooling curves, and red from warming curves. Comparing with the bulk, the temperature range of the phase transition shifts and expands to 180K-50K for the $1T' \rightarrow T_d$ transition, and 230K-320K for the $T_d \rightarrow 1T'$ transition. The expansion of $T_{hysteresis}$ can also be explained by the Barkhausen effect, as defects pinning impedes the phase transition process, leading to super-cooled /super-heated states. We notice that the total-height of the Barkhausen jumps in cooling is much larger than in warming. One possible explanation is that the pinning effect is stronger as thermal fluctuation is smaller in lower temperature, thus larger Barkhausen jumps are recorded during cooling than during warming.

To study the evolution of the phase transition temperature as a function of sample thickness, we fabricated several devices with thickness ranging from 6nm to 120nm. For samples between 20nm-100nm, the Barkhausen effect can be observed. We can define the "Barkhausen zone" as the range of temperature between the highest and the lowest kink temperature. For samples thicker than 100nm, Barkhausen jumps disappear and the $\rho(T)$ curves show bulk-like behavior with a shift of the transition temperature (see supplementary materials figure S2). Figure 3 plots the temperature ranges of the Barkhausen jumps versus sample thickness together with the phase transition temperature range for bulk crystals. The temperature ranges are plotted by dash bars for each thickness. We find that both the upper bound and the lower bound of Barkhausen zone extend approximately linearly when sample thickness is reduced. This phenomenon infers that the pinning effect enhances linearly when sample's layer number decreases, suggesting a possible surface origin of the pinning defects. No transition signal is observed for samples at or below 6nm (see supplementary materials figure S4). Since $1T'$ and $T_d$ MoTe$_2$ share the same in-plane structure[14], the phase transition behavior towards the monolayer limit warrants further investigation.

Figure 4 plots the histogram of the normalized resistivity kink height ($\Delta R/R$) for all the devices measured. Here $\Delta R = R_{1T'} - R_{T_d}$ represents ~5% of the total resistivity of the sample, thus we can assign the total resistivity of the sample $R = R_{1T'} \cong R_{T_d}$. We found that the distribution function $f$ of $\Delta R/R$ from all the devices we measured follows a power law $f \sim (\Delta R/R)^{-\alpha}$, with $= 1.27 \pm 0.08$. Theoretically, it has been predicted that the domain size ($s$) of each Barkhausen jump has a distribution $P$ that follows a power law $P(s) \sim s^{-\beta}$[22, 23, 28]. Mean-field theory gives a universal exponent $\beta = 2 - 2/(d + 1)$ for such power law behavior, where $d \leq 3$ is the dimension of the system[23]. It is reasonable to assume that $\Delta R$ is proportional to the total resistivity $R$ for a given size of flipped domain at any given temperature, thus the size of the flipped domain $s \sim \Delta R/R$. If the Barkhausen effect in thin MoTe$_2$ is three-dimensional, then $\beta = 1.5$, which does not agree with our experimental observation. On the other hand, if there are mainly two-dimensional phase pinning/depinning in thin MoTe$_2$, then $\beta = 4/3 \approx 1.33$, which agrees well with the experimentally extracted value of $\alpha$. This result suggests that the phase transition between $1T'$ and $T_d$ in thin MoTe$_2$ could be a rare example for non-magnetic Barkhausen effect in the two dimensional universality class. It is worth noting that the theories describing scaling behavior in a Barkhausen process in magnetic materials is far from united, much less is known for Barkhausen process in non-magnetic structural phase transition. For example, an elastic interface theory predicts

$\beta \approx 1.3$ for a three dimensional Barkhausen process[23, 36, 37]. Thus more theoretical study is needed to fully understand our experimental observation.

As a further verification of the shifting and expansion of the phase transition temperature, we performed temperature-dependent polarized Raman spectroscopy on thin MoTe$_2$ samples. The strongest Raman signal that distinguishes the 1$T'$ and the $T_d$ phases is the interlayer shear mode at ~13cm$^{-1}$, as measured in the parallel-polarized configuration [14]. Centrosymmetry breaking in the $T_d$ structure leads to a Raman active mode at 13cm$^{-1}$(denoted as the A peak), whereas this mode is Raman inactive in 1$T'$ MoTe$_2$. Figure 5a shows the Raman spectra between 6cm$^{-1}$ and 20cm$^{-1}$ of a 22nm-thick MoTe$_2$ sample at different temperature during sample cooling (left panel) and warming (right panel). The temperature-dependent evolution of the intensity of the A peak is shown in figure 5b. One can find that the intensity of the peak increases between 235K and 130K during sample cooling, and starts to decrease above 280K during sample warming. Up to 350K, the A peak is still observable for the 22nm sample. This is in stark contrast to bulk MoTe$_2$, where such change in the intensity of the A peak is completed within 215K to 280K[14]. Figure 5c is the temperature-dependent Raman spectra of a 155nm-thick MoTe$_2$ sample between 6cm$^{-1}$ and 20cm$^{-1}$, which shows intermediate behavior between bulk and the 22nm-thick MoTe$_2$. The observation in figure 5 is consistent with resistivity measurement, showing enhancing $T_{hysteresis}$ with samples of reducing thickness.

## Conclusions

We report the first observation of the Barkhausen effect in the first order structural phase transition between the two metallic phases, e.g., the 1$T'$ phase and the $T_d$ phase, of MoTe$_2$ thin flakes. The temperature range of the Barkhausen zone increase linearly with reducing thickness, pointing to a surface origin of the phase pinning defects. The distribution of normalized resistivity jumps ($\Delta R/R$) has a power law dependence with exponent $\alpha = 1.27 \pm 0.08$, suggesting an underlying scaling behavior. Temperature-dependent Raman spectroscopy also detects thickness dependent temperature ranges for the phase transition, consistent with data from transport measurement.

## Methods

The bulk 1$T'$ MoTe$_2$ crystals were synthesized by chemical vapor transport method [15, 38-42], and few-layer samples were mechanical exfoliated from bulk crystals and deposited onto silicon wafer with 300nm SiO$_2$. Standard electron beam lithography and metallization processes were performed to pattern multiple electrodes on few-layer MoTe$_2$ devices for four-probe measurements. The electrodes were made of 5nm Cr/ 50nm Au. Resistivity measurements were performed in PPMS with lock-in amplifiers. Speed of temperature sweep in the resistivity measurement is 5K/min. Higher and lower sweeping speeds are performed to confirm full thermalization of the sample during the measurement. Raman measurements were performed on exfoliated few-layer MoTe$_2$ in a confocal back-scattering geometry using a Horiba Jobin Yvon LabRAM HR Evolution spectrometer, equipped with 1800 gr/mm gratings and a liquid-nitrogen-cooled CCD. We used the $\lambda = 514$ nm line of an Argon laser for excitation. The BragGrate notch filters allow for measurements at low wave numbers. Polarized Raman spectra were measured in a-a configuration.

## Acknowledgments

This project has been supported by the National Basic Research Program of China (Grant Nos. 2014CB920900，2018YFA0305604), and the National Natural Science Foundation of China (Grant Nos.11774010 (C. Cao, X. Liu and J.-H. Chen)) (Grant Nos. 11674013,11704012 (D. Sun)). Crystal drawings were produced by VESTA[43,44].

Figure 1

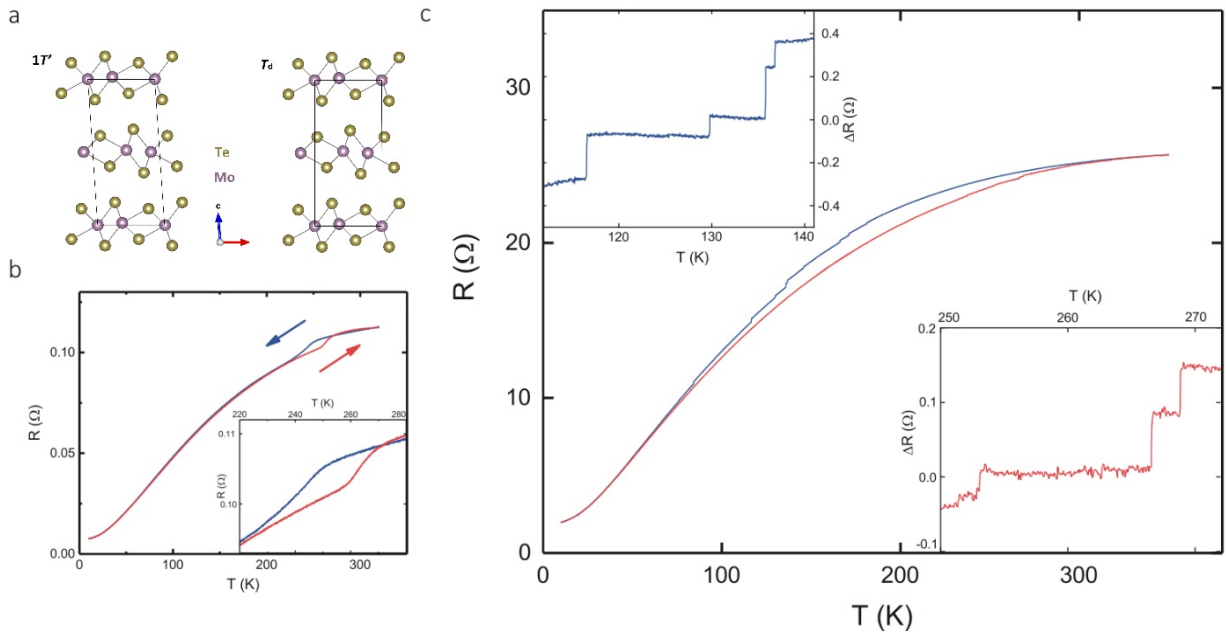

Fig1: **Temperature-induced phase transition in few-layer MoTe$_2$.**
(a) Crystal structures of $1T'$ and $T_d$ phase MoTe$_2$.
(b) Temperature dependent resistivity of a bulk MoTe$_2$ sample; cooling curve in blue, warming in red.
(c) Temperature dependent resistivity of a 50nm MoTe$_2$ sample. Cooling curve in blue, warming in red. Upper (lower) insets: zoom-in of the cooling (warming) curve in (c) with background subtracted.

Figure 2

a

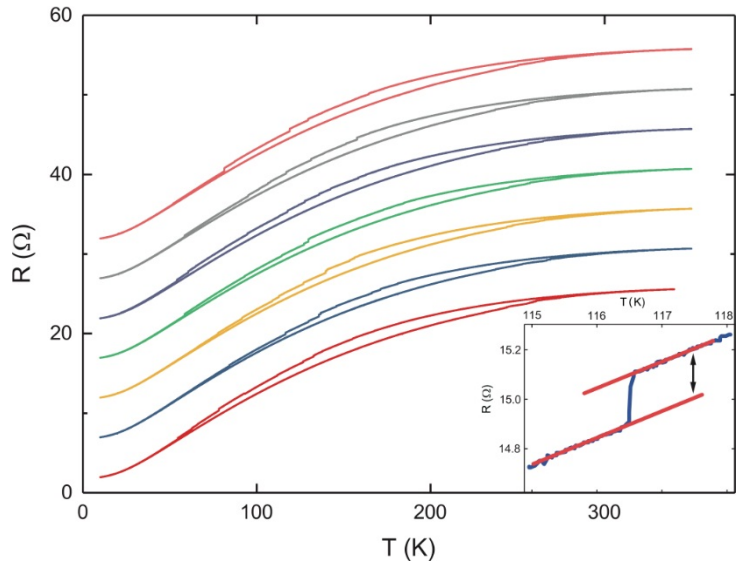

b

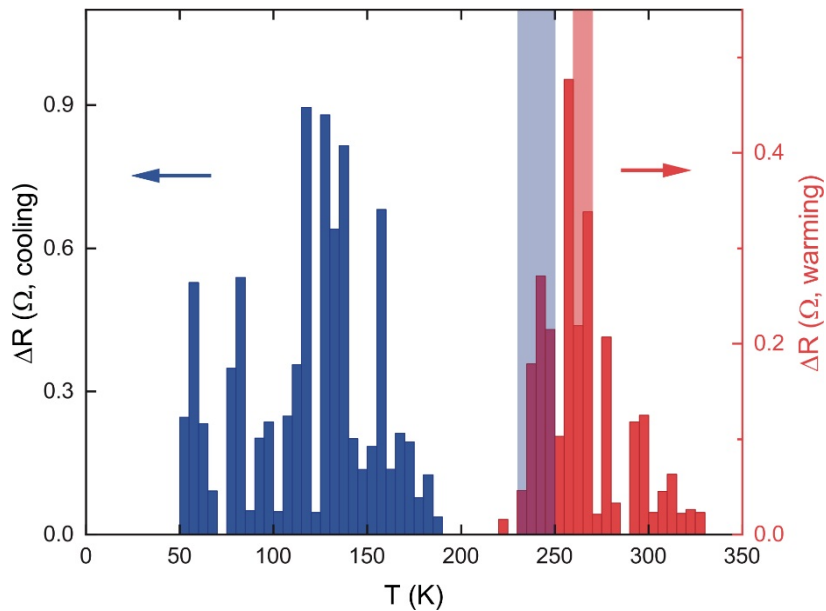

Fig2: **Barkhausen jumps in few-layer MoTe$_2$.**
(a) Repeated temperature dependent resistivity measurement of the 50nm MoTe$_2$ sample. Seven runs of the cooling-warming process are plotted together with Y-offset. The inset shows the method to determine kink height.
(b) Histogram of resistivity kinks in figure 2a, weighted by kink's height. Kinks in cooling and warming are plotted side-by-side. Blue- and red-shaded area is the phase transition region for bulk MoTe$_2$ crystals during cooling and warming, respectively.

Figure 3

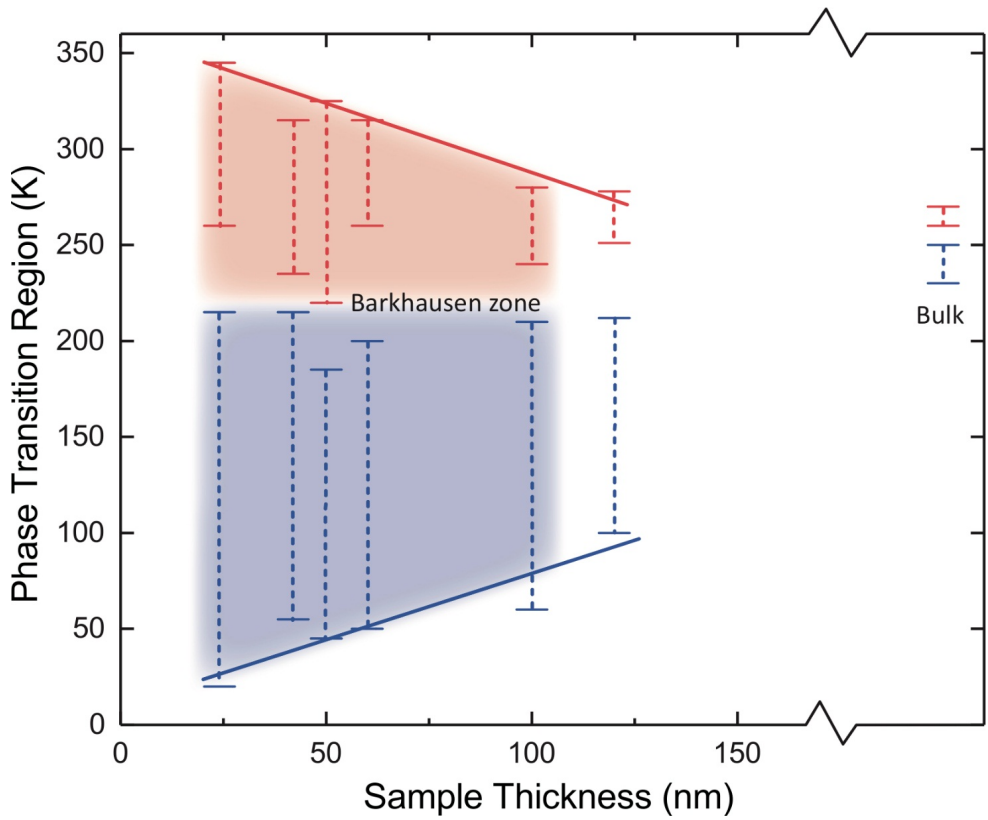

Fig3: **Thickness dependent phase transition temperature ranges.**
Barkhausen zone as a function of sample thickness. Barkhausen zone is extracted by the highest/lowest temperature that Barkhausen jumps appear, and are plotted by blue bars and red bars for cooling and warming, respectively. Shaded area denotes the parameter space of temperature and sample thickness where Barkhausen effect is detected. For sample thicker than 100nm, the endpoints are determined by the temperature at which cooling and warming curves begins to overlap.

Figure 4

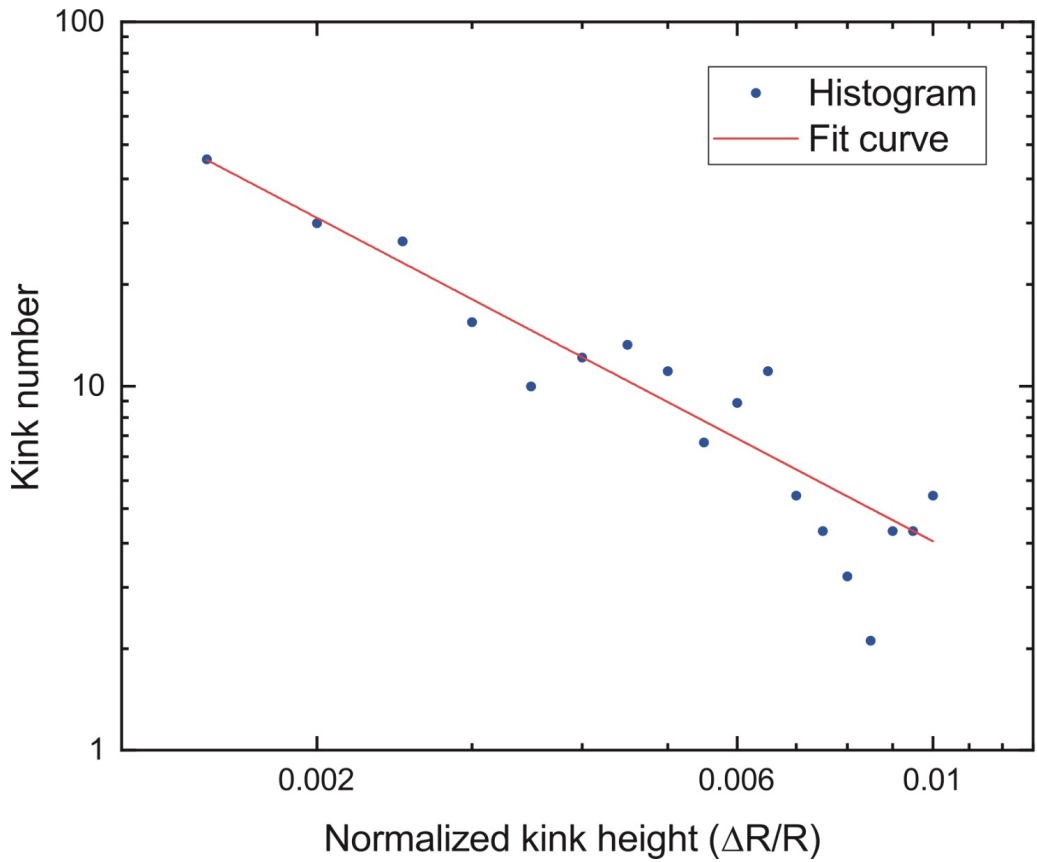

Fig4: **Distribution of normalized resistivity kink height.**
Histogram of normalized resistivity kink height $\Delta R/R$ for all the samples (blue dots), where $R$ is the sample resistivity at the resistivity jump, and $\Delta R$ is the respective jumps in resistivity. Red line shows a power law fitting. Exponent $\alpha = 1.27 \pm 0.08$ is obtained from the fit.

Figure5

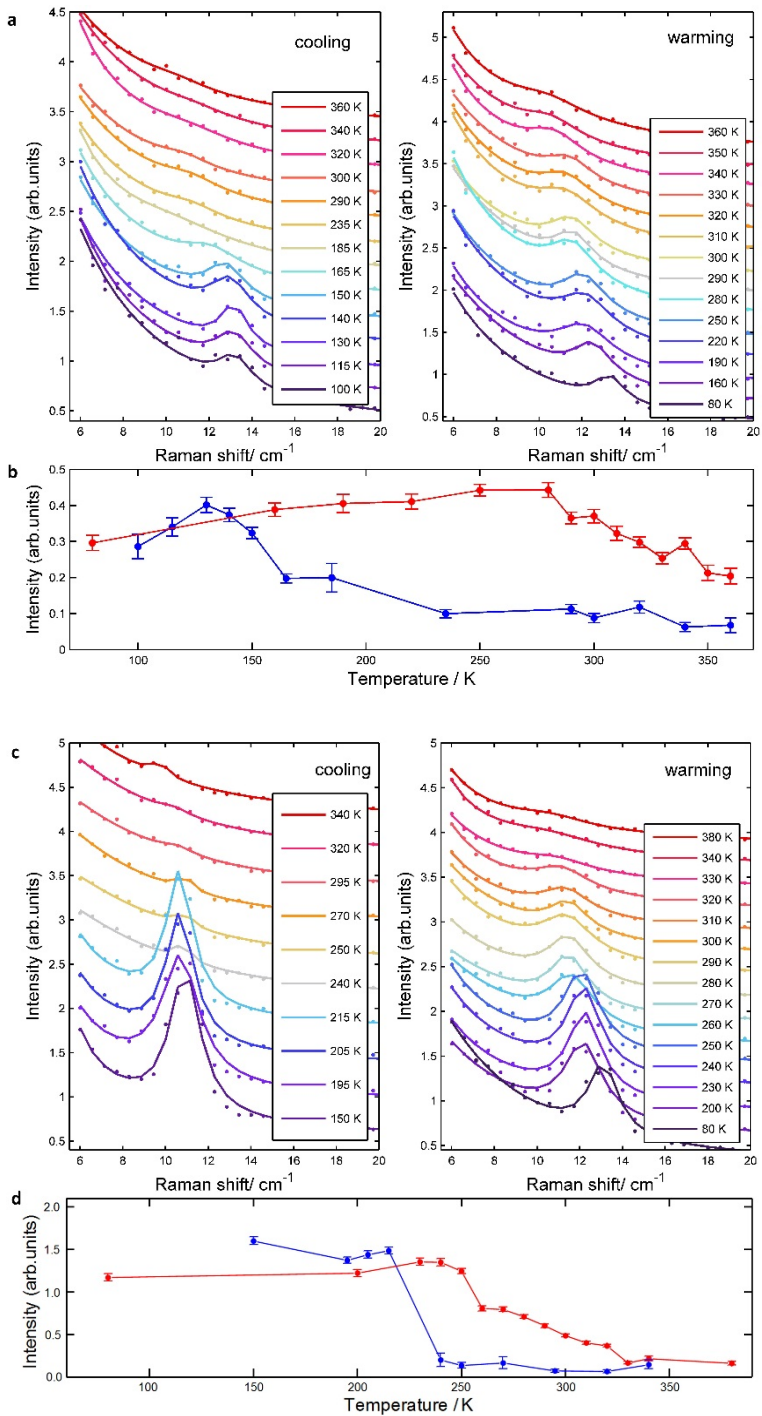

Fig5: **Temperature dependence of polarized Raman spectra for the signal at 13cm$^{-1}$.**
Temperature dependent Raman spectra around 13 cm$^{-1}$ for (a) a 22nm sample, (c) a 155 nm sample. Temperature dependence of the Raman intensity for the 13cm$^{-1}$ peak for (b) a 22nm sample, (d) a 155 nm sample. Blue for cooling and red for warming in (b) and (d).

# Supplementary materials

Figure.S1 to Figure.S5

Figure S1

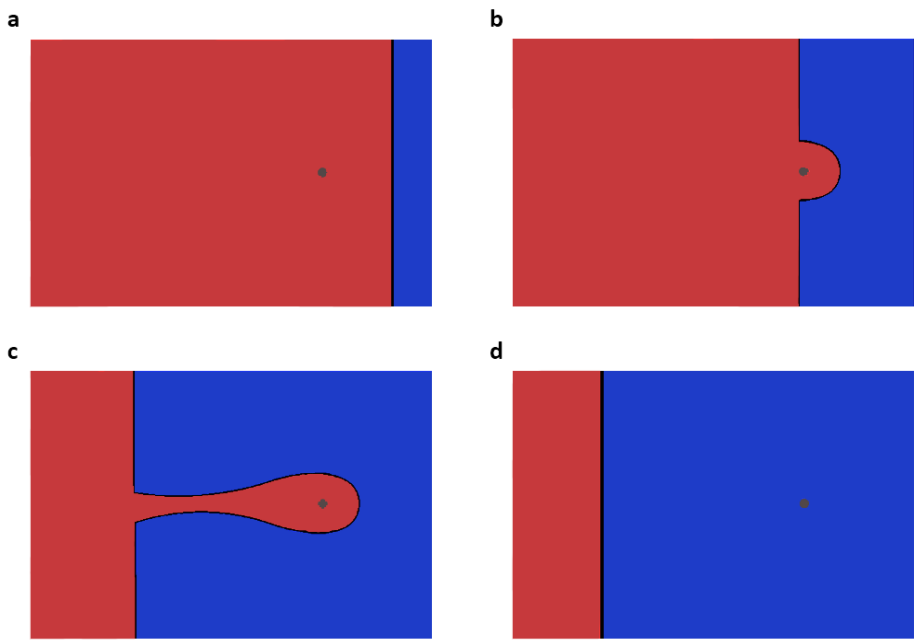

Fig.S1: **2D Schematic for the Barkhausen effect.**
Phase transition with Barkhausen effect is shown in time order: (a)→ (b) → (c) → (d). Blue and red area stands for two different states. The pinning defect is plotted as a gray dot. During the transition, the domain wall move through the defect, which pins its neighbor area in the past state (b,c). When de-pinning happens (d), the pinned area suddenly turns into the other state, which can be observed as a Barkhausen jump.

Figure S2

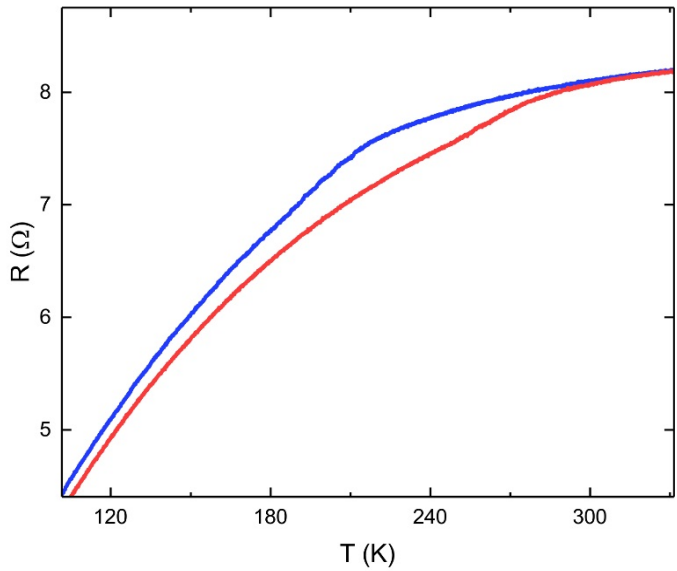

Fig.S2: **Resistivity of few-layer MoTe$_2$ out of the Barkhausen zone.**
Four-probe resistivity measurement of a 120nm MoTe$_2$ sample. Barkhausen jump disappear and normal phase transition curve remains. The transition appears at 200K and 250K for cooling and warming, respectively. Phase transition temperate is different from bulk behavior as shown in figure 1b.

Figure S3

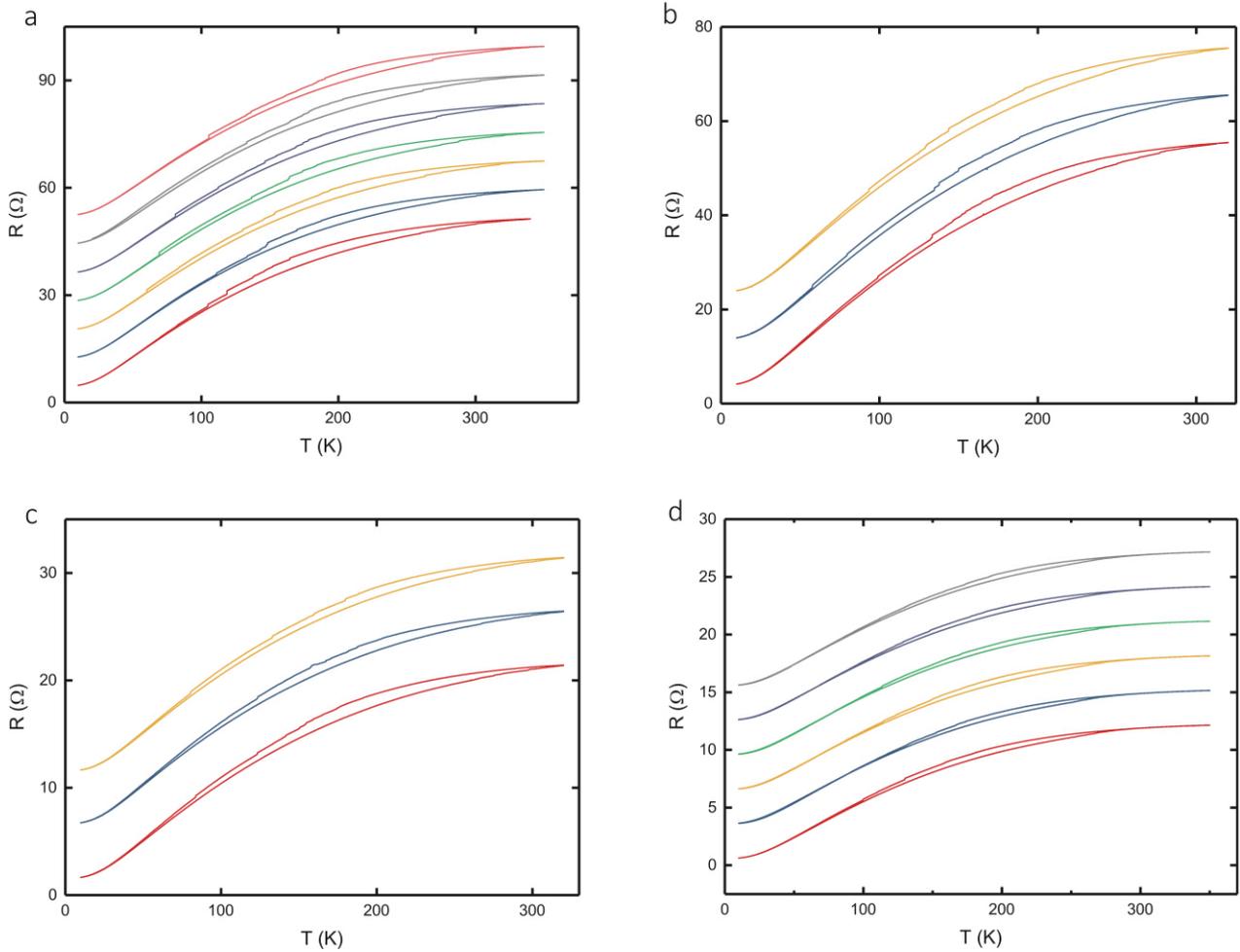

Fig.S3: **Barkhausen jumps in various few-layer MoTe$_2$ devices.**
Four-probe resistivity measurement of few-layer MoTe$_2$ samples with different thickness. Several runs of cooling-warming curves are plotted together with Y-offset. All the curves show resistivity kinks in a large temperature range. Sample thickness are (a) 24nm, (b) 42nm, (c) 60nm, and (d) 100nm, respectively.

Figure S4

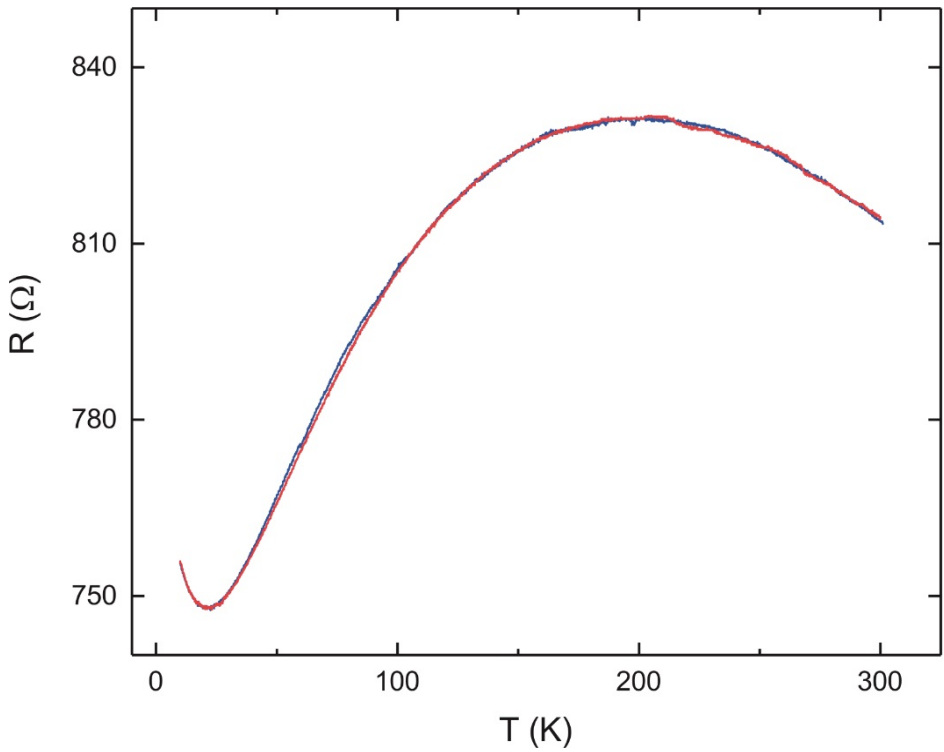

Fig.S4: **Resistivity of a 6nm MoTe$_2$ device.**
Four-probe resistivity measurement of a 6nm MoTe$_2$ sample. The cooling (blue) and warming (red) curve overlap from 300K to 10K. No structural phase transition signal is observed from electrical transport measurement.

Figure S5

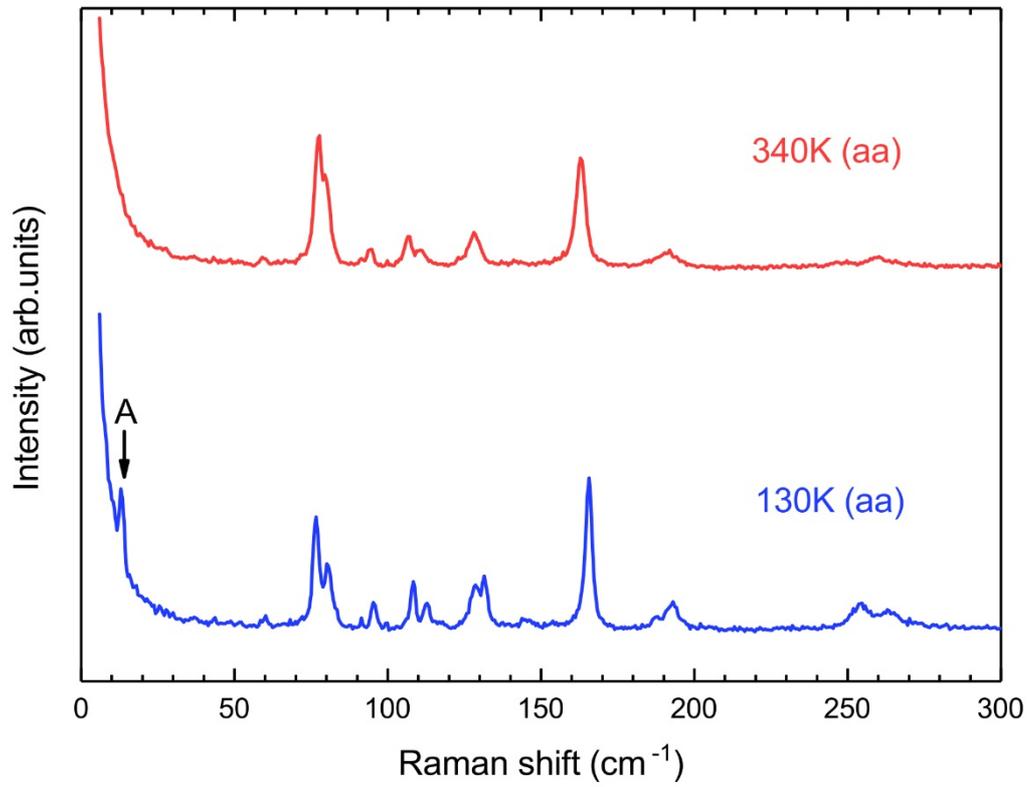

Fig.S5: **Full Raman spectra of few-layer MoTe$_2$.**
Polarized Raman spectra of a 22nm sample in aa configuration. The red curve was measured at 340K during cooling. A peak is unobservable and the sample stayed at 1T' phase. The blue curve was measured at 130K during cooling. The A peak emerged as the sample transformed to the $T_d$ phase.